\documentclass[epj]{svjour}

\usepackage{graphicx}
\usepackage{fancyhdr}
\def\makeheadbox{{
 }}
\def\mytitle{My title} 
\def\myauthors{My name}  
\def\mytype{My type of session}
\def\mysession{My session}

\def\mytitle{Soft-gluon resummation for Higgs differential distributions at the Large Hadron Collider}
\def\myauthors{Giuseppe Bozzi}
\def\mytype{Contributed Talk}
\def\mysession{Colliders - Higgs Phenomenology}

\begin{document}

\title{Soft-gluon resummation for Higgs differential distributions at the Large Hadron Collider}
\author{Giuseppe Bozzi\inst{1}
\thanks{\emph{Email:} giuseppe@particle.uni-karlsruhe.de}} 
\institute{Institut f\"ur Theoretische Physik, Universit\"at Karlsruhe, P.O.Box 6980, 76128 Karlsruhe, Germany}
\date{}
\abstract{We study the transverse-momentum ($q_T$) and rapidity ($y$) distributions of the Higgs boson in perturbative QCD, including the most advanced theoretical information presently available: fixed-order perturbation theory at Next-to-Leading Order (NLO) in the large-$q_T$ region ($q_T \sim M_H$, being $M_H$ the Higgs mass), and soft-gluon resummation at the Next-to-Next-to-Leading Logarithmic accuracy (NNLL) in the small-$q_T$ region ($q_T \ll M_H$). We present numerical results for the doubly-differential ($q_T$ and $y$) cross section for the production of a Standard Model Higgs boson at the Large Hadron Collider (LHC).\PACS{{12.38.Bx}{Perturbative calculations} \and {12.38.Cy}{Summation of perturbation theory} \and {14.80.Bn}{Standard-model Higgs bosons}}} 

\maketitle

\section*{Introduction}
The gluon fusion process $gg \to H$, which proceeds through a heavy-quark loop, is the main production mechanism for the Standard Model Higgs boson \cite{Hrev} at the Large Hadron Collider (LHC) over the full mass range 100 GeV $\leq M_H \leq$ 1 TeV \cite{atlascms}. As a consequence, in the last decade an enormous theoretical effort has been devoted to the computation of higher-order perturbative corrections both for this signal and for its main backgrounds, in order to achieve the highest possible theoretical accuracy. 

A particularly important observable is the doubly-differential transverse-momentum ($q_T$) and rapidity ($y$) distribution: a precise knowledge of the Higgs $q_T$- and $y$-spectrum is very important to improve the statistical significance at hadron colliders by applying, for instance, suitable cuts on the jets accompanying the Higgs decay products \cite{atlascms,Carena:2000yx}. 

At ${\cal O} (\alpha_S^2)$, the Higgs is produced with vanishing transverse-momentum. In order to have non-vanishing values of $q_T$, a recoiling jet is required and thus the Leading Order (LO) transverse-momentum distribution starts at ${\cal O} (\alpha_S^3)$ \cite{Ellis:1987xu}. The Next-to-Leading Order (NLO) QCD corrections have been calculated in the infinite top mass limit ($M_t \to \infty$) \cite{deFlorian:1999zd,Ravindran:2002dc,Glosser:2002gm,Anastasiou:2005qj,Catani:2007vq}, i.e. by using an effective lagrangian directly coupling the Higgs to gluons. This approximation has proved to be sufficiently accurate provided that $M_H \leq 2M_t$ and $q_T \leq M_t$ \cite{DelDuca:2001fn,Smith:2005yq}.

It has long been known that, in the small-$q_T$ region ($q_T \ll M_H$), the presence of large logarithmic terms of the form $\alpha_S \log (M_H^2/q_T^2)$ spoils the convergence of the perturbative series. These terms originate from the emission of soft and collinear radiation from the incoming partons. Since the bulk of the events is expected in the small-$q_T$ region, an all order summation of the logarithmic enhancements is mandatory in order to obtain reliable results.

The technique to perform soft-gluon resummation at small transverse-momentum in perturbative QCD is well-known \cite{Dokshitzer:hw,Parisi:1979se,Curci:1979bg,Collins:1981uk,Kodaira:1981nh,Collins:1984kg,Catani:vd,Catani:2000vq} and has been applied to the Higgs case up to Next-to-Next-to-Leading Logarithmic level (NNLL) \cite{deFlorian:2000pr}. The fixed-order and resummed results have eventually to be matched in order to prevent possible double-counting of the logarithmic terms in the intermediate-$q_T$ region and thus to obtain a uniform theoretical accuracy over the entire $q_T$-range. The mat-ching is achieved by taking the sum of the two contributions and then subtracting the truncation of the resummed term to the same perturbative order of the fixed-order result.

In \cite{Bozzi:2003jy,Bozzi:2005wk} we provided the details of the transverse-momentum resummation formalism that we developed for the hadroproduction of a general colourless final state, and we performed a detailed phenomenological study in the case of Higgs boson production at the LHC. We included the NNLL resummed result and the purely perturbative calculation at NLO, thus reaching a uniform theoretical accuracy of ${\cal O} (\alpha_S^4)$ over the entire $q_T$-range. The formalism has been implemented in the publicly available numerical code {\texttt HqT} \cite{HqT}. Lately \cite{Bozzi:2007pn} we extended the resummation formalism to include rapidity dependence, thus providing NNLL+NLO accuracy for the {\it fully-differential} cross section in $q_T$ and $y$ at the LHC. 
\newpage
The inclusion of rapidity does not change the main features of our formalism:
\begin{itemize}
\item the resummation is performed at the level of the partonic cross section with factorization of the parton distribution functions as in the customary fixed-order calculations;
\item the formalism can be applied to any hard-scattering process producing a colourless final state accompanied by an arbitrary and undetected final state;
\item the singular terms are exponentiated in a {\it universal} (i.e. process-independent) form factor;
\item a constraint of perturbative unitarity imposed on the resummed contribution allows both to decrease the uncertainty in the matching procedure at intermediate $q_T$ and to recover the total cross section result upon integration over $q_T$. 
\end{itemize}

In the following we will show numerical results for the Higgs differential distributions at the LHC. For details about the formalism and for further phenomenological discussions, we refer the reader to our previous papers \cite{Bozzi:2003jy,Bozzi:2005wk,Bozzi:2007pn}. 

\section*{Numerical results}

We present numerical results for the doubly-differential ($q_T$ and $y$) cross section for the production of a Standard Model Higgs boson with mass $M_H$=125 GeV at the Large Hadron Collider. We used the MRST2004 NNLO (NLO) set of parton distribution functions \cite{Martin:2004ir} with $\alpha_S$ evaluated at 3 loops (2 loops) for predictions at NNLL+NLO (NLL+LO) accuracy. We fixed the renormalization and factorization scales both equal to the Higgs mass $\mu_R=\mu_F=M_H$ and made them vary between $M_H/2$ and $2M_H$ to examine the scale dependence of our results. As a cross-check of our calculation, we have verified that we reobtain both the numerical results of Ref.\cite{Bozzi:2005wk} upon integration over $y$ and the NNLO total cross section at fixed $y$ \cite{Anastasiou:2005qj,Catani:2007vq} upon integration over $q_T$.

\begin{figure}
\includegraphics[width=0.45\textwidth,height=0.45\textwidth,angle=0]{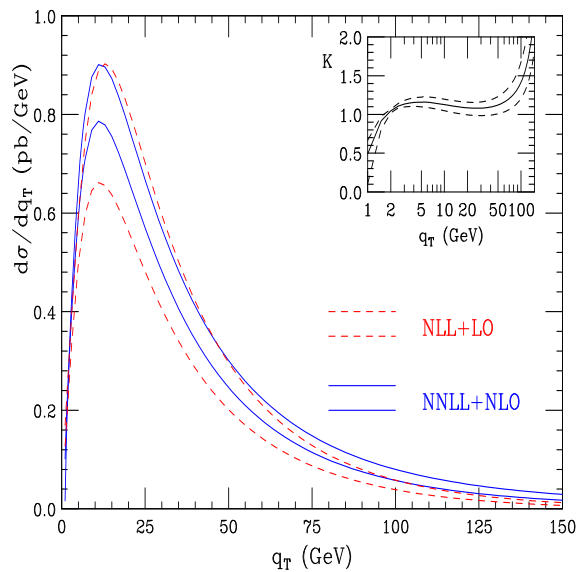}
\caption{Comparison of the NLL+LO and NNLL+NLO bands. The NNLL+NLO result normalized to the central value of the NLL+LO result (see Eq.~\ref{eq:1}) is shown in the inset plot (from \cite{Bozzi:2005wk}).}
\label{fig:1}
\end{figure}

In Fig.~\ref{fig:1} the scale dependence of the NLL+LO and NNLL+NLO $q_T$-spectrum with integrated rapidity is shown: the reduced thickness of the NNLL+NLO band with respect to the NLL+LO one and the overlapping of the two bands in the region $q_T\leq$ 100 GeV indicate a very good convergence of the resummed result. In the upper-right corner we show the $K$ factor defined by
\begin{equation}
\label{eq:1}
K(q_T)=\frac{d\sigma_{NNLL+NLO}(\mu_F,\mu_R)}{d\sigma_{NLL+LO}(\mu_F=\mu_R=M_H)},
\end{equation}
i.e., the NNLL+NLO band normalized to the central value of the NLL+LO one. We note that a simple rescaling of the NLL+LO result is not allowed since the $K$ factor turns out to be $q_T$-dependent: the cross section is enhanced in the large-$q_T$ region, where higher-order contributions are not negligible \cite{deFlorian:1999zd,Ravindran:2002dc,Glosser:2002gm}, and suppressed in the small-$q_T$ region, where the non-per-turbative regime sets in.

\begin{figure}
\includegraphics[width=0.45\textwidth,height=0.45\textwidth,angle=0]{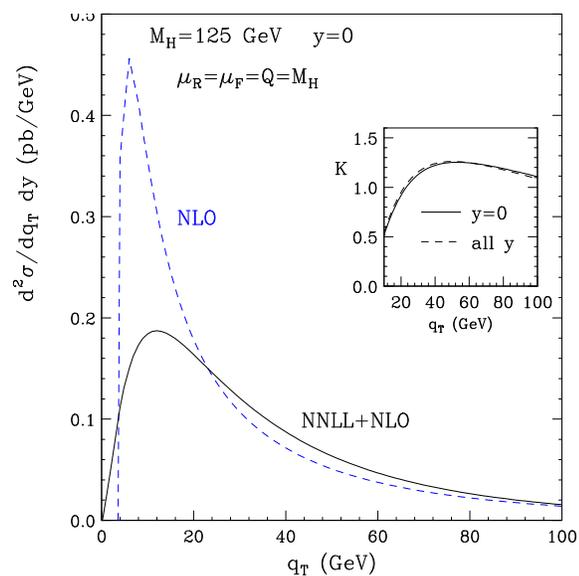}
\caption{The Higgs $q_T$-spectrum at the LHC for $y$=0. The $K$ factor defined in Eq.~\ref{eq:2} is shown in the inset plot (from \cite{Bozzi:2007pn}).}
\label{fig:2}
\end{figure}

In Fig.~\ref{fig:2} we plot the $q_T$ dependence of the cross section at $y$=0, showing both the purely perturbative NLO result and the resummed NNLL+NLO result. The NLO cross section diverges to $-\infty$ as $q_T\to$ 0 due to the large logarithmic terms coming from soft-gluon radiation, and shows an unphysical peak. In contrast, the NNLL+NLO result is perfectly regular at small-$q_T$, vanishing for $q_T$=0 and converging to the NLO result for higher $q_T$ values ($q_T \sim M_H$). The resummation effects are clearly visible when looking at the inside plot, where the ratio of the matched NNLL+NLO result to the NLO fixed-order result is shown:

\begin{equation}
\label{eq:2}
K(q_T,y)=\frac{d\sigma_{NNLL+NLO}/(dq_T \, dy)}{d\sigma_{NLO}/(dq_T \, dy)}.
\end{equation}

Resummation is not only relevant at small $q_T$ but also in the intermediate region ($q_T\leq$80 GeV), where there is a $\sim$20\% enhancement with respect to fixed-order. The small difference between the $y$=0 curve (solid line) and the integrated rapidity result (dashed line) evidentiates the poor rapidity dependence of the resummed result.

\begin{figure}
\includegraphics[width=0.45\textwidth,height=0.45\textwidth,angle=0]{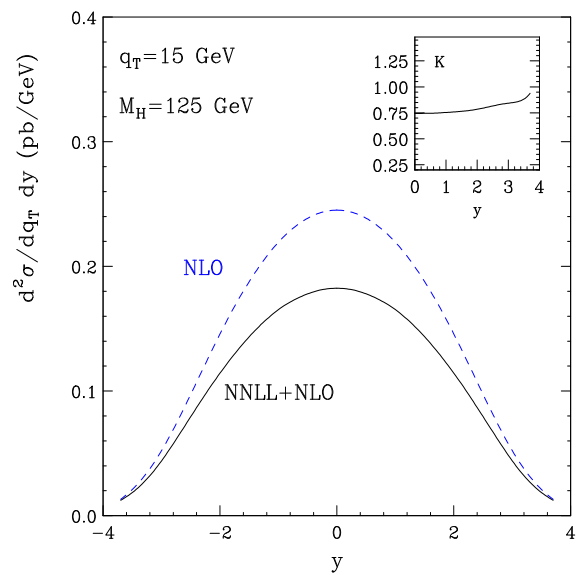}
\caption{The Higgs $y$-spectrum at the LHC for $q_T$=15 GeV. The $K$ factor defined in Eq.~\ref{eq:2} is shown in the inset plot (from \cite{Bozzi:2007pn}).}
\label{fig:3}
\end{figure}

In Figure~\ref{fig:3} the rapidity dependence of the cross section at $q_T$=15 GeV is shown both at NLO (dashes) and NNLL+NLO (solid line) accuracy. The resummed result reduces the cross section in the central rapidity region, where most of the events are expected ($\sim$25\% suppression with respect to fixed-order). The $K$ factor of Eq.~\ref{eq:2}, shown in the inset plot, is roughly constant in the central rapidity region and starts to be rapidity-dependent in the forward (and backward) region where the cross section is rather small. This behaviour also explains the coincidence of the two curves in the inset plot of Fig.~\ref{fig:2}.

A more detailed investigation of the rapidity dependence of the cross section is obtained by studying the quantity
\begin{equation}
\label{eq:3}
R(q_T;y)=\frac{d^2\sigma/(dq_T \,dy)}{d\sigma/dq_T}
\end{equation}
and its $q_T$-integrated version
\begin{equation}
\label{eq:4}
R_y=\frac{d\sigma/dy}{\sigma}. 
\end{equation}

\begin{figure}
\includegraphics[width=0.45\textwidth,height=0.45\textwidth,angle=0]{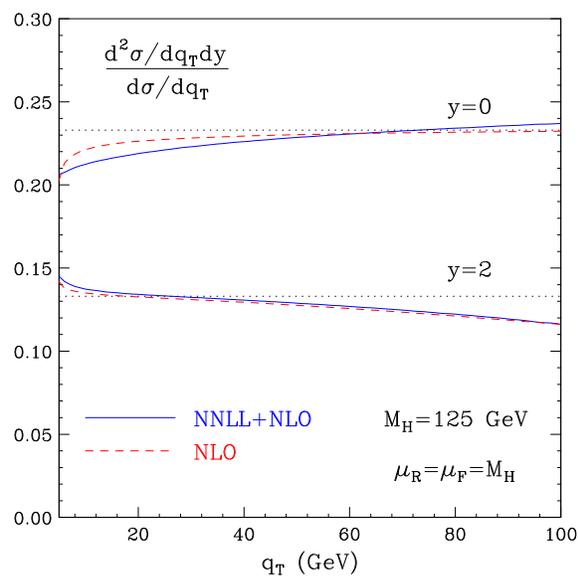}
\caption{The Higgs rescaled $q_T$-spectrum at the LHC as defined in Eq.~\ref{eq:3}. The dotted lines show the values for the ratio $R_y$ defined in Eq.~\ref{eq:4} (from \cite{Bozzi:2007pn}).}
\label{fig:4}
\end{figure}

In Fig.~\ref{fig:4} we plot these two quantities, as a function of $q_T$, for the two different values $y$=0 and $y$=2. The NNLL+NLO and NLO results are nearly equal at fixed rapidity, reflecting the similar behaviour of the inset plot in Fig.~\ref{fig:2}. The overall decrease of the differential cross section when going from $y$=0 to $y$=2 amounts to nearly 40\%, as expected since the total cross section rapidly decreases with increasing rapidity. As for the $q_T$ dependence, the results show a slightly increasing (decreasing) slope for $y$=0 ($y$=2) and it is quite evident that the cross section varies more in absolute value than in $q_T$ shape. 

\section*{Summary}

We applied the resummation formalism to the fully-differential cross section for the production of the Higgs boson at the LHC, combining the most advanced perturbative information available at present: soft-gluon resummation at NNLL accuracy and fixed-order prediction at NLO QCD. The numerical results show a sizeable resummation effect at intermediate $q_T$, a $q_T$-shape mildly dependent on the rapidity of the Higgs boson and an overall stability of the cross section with respect to scale variation and to the inclusion of higher perturbative orders.

{\bf Acknowledgements.} This work was supported by the Deutsche Forschungsgemeinschaft under SFB TR-9 ``Computergest\"utzte Theoretische Teilchenphysik''.


\begin{thebibliography}{999}

\bibitem{Hrev}
For a review on Higgs physics in and beyond the Standard Model, see J.~F.~Gunion, H.~E.~Haber, G.~L.~Kane and S.~Dawson, {\it The Higgs Hunter's Guide} (Addison-Wesley, Reading, Mass., 1990); M.~Carena and H.~E.~Haber, Prog.\ Part.\ Nucl.\ Phys.\  {\bf 50} (2003) 63; A.~Djouadi, report LPT-ORSAY-05-17 [hep-ph/0503172], report LPT-ORSAY-05-18 [hep-ph/0503173].

\bibitem{atlascms}
ATLAS Coll., {\it ATLAS Detector and Physics Performance: Technical Design Report}, Vol. 2, report CERN/LHCC/99-15 (1999); S.~Asai {\it et al.}, Eur.\ Phys.\ J.\ C {\bf 32S2} (2004) 19; CMS Coll.,{\it  CMS Physics Technical Design Report: Physics Performance}, Vol. 2, report CERN/LHCC/2006-021 (2006).

\bibitem{Carena:2000yx}
M.~Carena {\it et al.}, {\it Report of the Tevatron Higgs working group}, hep-ph/0010338; CDF and D0 Collaborations, {\it Results of the Tevatron Higgs Sensitivity Study}, report FERMILAB--PUB--03/320-E; V.~M.~Abazov {\it et al.}  [D0 Coll.], Phys.\ Rev.\ Lett.\  {\bf 96} (2006) 011801; A.~Abulencia {\it et al.}  [CDF Coll.], Phys.\ Rev.\ Lett.\  {\bf 97} (2006) 081802; The TEVNPH working group [for the CDF and D0 Collaborations], hep-ex/0612044.

\bibitem{Ellis:1987xu}
R.~K.~Ellis, I.~Hinchliffe, M.~Soldate and J.~J.~van der Bij, Nucl.\ Phys.\ B {\bf 297} (1988) 221; U.~Baur and E.~W.~Glover, Nucl.\ Phys.\ B {\bf 339} (1990) 38.

\bibitem{deFlorian:1999zd}
D.~de Florian, M.~Grazzini and Z.~Kunszt, Phys.\ Rev.\ Lett.\  {\bf 82} (1999) 5209.

\bibitem{Ravindran:2002dc}
V.~Ravindran, J.~Smith and W.~L.~Van Neerven, Nucl.\ Phys.\ B {\bf 634} (2002) 247.

\bibitem{Glosser:2002gm}
C.~J.~Glosser and C.~R.~Schmidt, JHEP {\bf 0212} (2002) 016.

\bibitem{DelDuca:2001fn}
V.~Del Duca, W.~Kilgore, C.~Oleari, C.~Schmidt and D.~Zeppenfeld, Nucl.\ Phys.\ B {\bf 616} (2001) 367, Phys.\ Rev.\ D {\bf 67} (2003) 073003.

\bibitem{Smith:2005yq}
J.~Smith and W.~L.~van Neerven, Nucl.\ Phys.\  B {\bf 720} (2005) 182.

\bibitem{Anastasiou:2005qj}
C.~Anastasiou, K.~Melnikov and F.~Petriello, Phys.\ Rev.\ Lett.\  {\bf 93} (2004) 262002, Nucl.\ Phys.\  B {\bf 724} (2005) 197.

\bibitem{Catani:2007vq}
S.~Catani and M.~Grazzini, Phys.\ Rev.\ Lett.\ {\bf 98} (2007) 222002.

\bibitem{Dokshitzer:hw}
Y.~L.~Dokshitzer, D.~Diakonov and S.~I.~Troian, Phys.\ Lett.\  B {\bf 79} (1978) 269, Phys.\ Rep.\  {\bf 58} (1980) 269.

\bibitem{Parisi:1979se}
G.~Parisi and R.~Petronzio, Nucl.\ Phys.\ B {\bf 154} (1979) 427.

\bibitem{Curci:1979bg}
G.~Curci, M.~Greco and Y.~Srivastava, Nucl.\ Phys.\ B {\bf 159} (1979) 451.

\bibitem{Collins:1981uk}
J.~C.~Collins and D.~E.~Soper, Nucl.\ Phys.\ B {\bf 193} (1981) 381 [Erratum-ibid.\ B {\bf 213} (1983) 545], Nucl.\ Phys.\ B {\bf 197} (1982) 446.

\bibitem{Kodaira:1981nh}
J.~Kodaira and L.~Trentadue, Phys.\ Lett.\ B {\bf 112} (1982) 66, report SLAC-PUB-2934 (1982), Phys.\ Lett.\ B {\bf 123} (1983) 335.

\bibitem{Collins:1984kg}
J.~C.~Collins, D.~E.~Soper and G.~Sterman, Nucl.\ Phys.\ B {\bf 250} (1985) 199.

\bibitem{Catani:vd}
S.~Catani, E.~D'Emilio and L.~Trentadue, Phys.\ Lett.\ B {\bf 211} (1988) 335.

\bibitem{Catani:2000vq}
S.~Catani, D.~de Florian and M.~Grazzini, Nucl.\ Phys.\ B {\bf 596} (2001) 299.

\bibitem{deFlorian:2000pr}
D.~de Florian and M.~Grazzini, Phys.\ Rev.\ Lett.\ {\bf 85} (2000) 4678, Nucl.\ Phys.\ B {\bf 616} (2001) 247.

\bibitem{Bozzi:2003jy}
G.~Bozzi, S.~Catani, D.~de Florian and M.~Grazzini, Phys.\ Lett.\ B {\bf 564} (2003) 65.

\bibitem{Bozzi:2005wk}
G.~Bozzi, S.~Catani, D.~de Florian and M.~Grazzini, Nucl.\ Phys.\ B {\bf 737} (2006) 73.

\bibitem{HqT}
{\texttt http://theory.fi.infn.it/grazzini/codes.html}

\bibitem{Bozzi:2007pn}
G.~Bozzi, S.~Catani, D.~de Florian and M.~Grazzini, arXiv:0705.3887 [hep-ph].

\bibitem{Martin:2004ir}
A.~D.~Martin, R.~G.~Roberts, W.~J.~Stirling and R.~S.~Thorne, Phys.\ Lett.\ B {\bf 604} (2004) 61.

\end{thebibliography}
\end{document}